\documentclass[sn-mathphys-num,iicol]{sn-jnl}

\usepackage{graphicx}%
\usepackage{multirow}%
\usepackage{amsmath,amssymb,amsfonts}%
\usepackage{amsthm}%
\usepackage{mathrsfs}%
\usepackage[title]{appendix}%
\usepackage{xcolor}%
\usepackage{textcomp}%
\usepackage{manyfoot}%
\usepackage{booktabs}%
\usepackage{algorithm}%
\usepackage{algorithmicx}%
\usepackage{algpseudocode}%
\usepackage{listings}%
\usepackage{aas_macros}

\theoremstyle{thmstyleone}%
\theoremstyle{thmstyletwo}%

\theoremstyle{thmstylethree}%

\newcommand{\ion}[2]{#1\,\textsc{#2}}
\newcommand{\as}{$^{\prime\prime}$}
\newcommand{\kms}{km\,s$^{-1}$}

\begin{document}

\title{Applications of Atomic Data to Studies of the Sun}


\author[1,2]{\fnm{Peter R.} \sur{Young}}\email{peter.r.young@nasa.gov}

\affil[1]{\orgdiv{Heliophysics Division}, \orgname{NASA Goddard Space Flight Center},  \city{Greenbelt}, \state{MD}, \postcode{20771},  \country{USA}}

\affil[2]{\orgdiv{Department of Mathematics, Physics \& Electrical Engineering}, \orgname{Northumbria University}, \city{Newcastle Upon Tyne}, \postcode{NE1 8ST}, \country{UK}}


\abstract{The Sun is a standard reference object for Astrophysics and also a fascinating subject of study in its own right. X-ray and extreme ultraviolet movies of the Sun's atmosphere show an extraordinary diversity of plasma phenomena, from barely visible bursts and jets to coronal mass ejections that impact a large portion of the solar surface. The processes that produce these phenomena, heat the corona and power the solar wind remain actively studied and accurate atomic data are essential for interpreting observations and making model predictions.  For the Sun's interior 
intense effort is focused on resolving the ``solar problem," (a discrepancy between solar interior models and helioseismology measurements) and atomic data are central to both element abundance measurements and interior physics such as opacity and nuclear reaction rates.
In this article, topics within solar interior and solar atmosphere physics are discussed and the role of atomic data described. Areas of active research are highlighted and specific atomic data needs are identified.
}


\maketitle

\section{Introduction}\label{sec1}

Solar Physics encompasses studies of the solar interior and the solar atmosphere. The latter extends far from the Sun and transitions to the solar wind, thus there is significant overlap with Heliospheric Physics particularly with regard the origin of the solar wind and how it is accelerated.  Atomic data are needed in order to model radiative emissions from the solar plasma and compare with remote sensing observations. They are also required for modeling the solar interior, in particular opacities and nuclear cross-sections. 
This article is based on an invited talk \cite{young_2024_10908729} given at the 14th Atomic Spectra and Oscillator Strengths conference\footnote{\url{https://asos2023paris.sciencesconf.org/}.} held in Paris in July 2023. It describes recent advances in applying atomic data to Solar Physics and highlights areas where new data are needed.

Solar Physics is a field that is strongly motivated by new observational data, particularly from space. The US space agency, NASA, dominates solar space observations as shown in Table~\ref{tbl:missions}. Fourteen current and future major space missions are listed, with six led by NASA and four having NASA as a major partner. The growing strength of China and India is reflected in the three recent missions from these countries.

Given the dominant impact of NASA on the field, it is appropriate to expand on the role of Solar Physics within NASA. The Science Mission Directorate (SMD) within NASA funds basic science, including space missions and research. There are four major science divisions within SMD: Earth Science, Planetary Science, Astrophysics and Heliophysics. Solar Physics falls within the latter, which also includes Heliospheric Physics (study of the solar wind), and physics of the near-Earth plasma environment (magnetosphere, ionosphere, thermosphere and mesosphere). One focus of Heliophysics is Space Weather, the effect of solar activity on the Earth, particularly its effect on astronauts and space- and ground-based infrastructure.

\begin{table}[t]
\caption{Major Solar Physics space missions. Acronyms are given in Table~\ref{tbl:acronyms}.}\label{tbl:missions}%
\begin{tabular}{@{}lll@{}}
\toprule
Mission & Sponsor & Launch\\
\midrule
SOHO & ESA\footnotemark[1] (Europe) & 1995 \\
STEREO & NASA (USA) & 2006 \\
Hinode & JAXA\footnotemark[1]  (Japan) & 2006 \\
SDO & NASA & 2010 \\
IRIS & NASA & 2014 \\
Parker Solar Probe & NASA & 2018 \\
Solar Orbiter & ESA\footnotemark[1]   & 2020 \\
CHASE & CNSA (China) & 2021 \\
ASO-S & CNSA  & 2022\\
Aditya-L1 & ISRO (India) & 2023 \\
PUNCH & NASA & 2025\footnotemark[2] \\
MUSE & NASA & 2027\footnotemark[2]\\
Solar-C & JAXA\footnotemark[1]  & 2028\footnotemark[2]\\
Vigil & ESA\footnotemark[1]  & 2031\footnotemark[2] \\
\botrule
\end{tabular}
\footnotetext[1]{NASA a major partner.}
\footnotetext[2]{Expected launch dates.}

\end{table}

A dedicated Division for Heliophysics ensures there is regular cadence of NASA Solar Physics missions. However, support for atomic data calculations and measurements within NASA Heliophysics is more limited than for NASA Astrophysics.  For Astrophysics and Solar Physics, atomic data are needed to interpret the radiative emissions measured by telescopes. 
Heliophysics includes studies of solar wind and Earth plasmas that can be directly sampled by instrumentation, hence atomic data are not required for interpreting these data. Thus demand for new atomic data within Heliophysics is mostly confined to the remote-sensing Solar Physics community. Of course there is a wide overlap between atomic data needs for Astrophysics and Solar Physics, so the latter community benefits from advances made for Astrophysics. 

Section~\ref{sec:interior} summarizes some properties of the solar interior and how it is studied, and Section~\ref{sec:int-appl} gives examples of how atomic data is applied to studies of the interior and photosphere. Properties of the solar atmosphere are summarized in Section~\ref{sec:atmosphere}, and applications of atomic data are described in Section~\ref{sec:atm-appl}. A summary is given in Section~\ref{sec:summary}. Acronyms for Solar Physics missions and instruments mentioned in this article are given in Table~\ref{tbl:acronyms}.

\section{The Solar Interior and Photosphere}\label{sec:interior}

The Sun's interior comprises the core where nuclear fusion occurs, the radiative zone where energy is transferred outwards by photons, and the convection zone where energy is transferred by convective motions. The boundaries between the regions are at 0.25\,$R_\odot$ and 0.70\,$R_\odot$. The solar rotation changes at the latter boundary (termed the tachocline) from uniform in the inner region to differential in the outer region, which is believed to play an important role in the solar dynamo. The solar photosphere---usually considered the surface of the Sun---is where the Sun transitions from being optically thick to radiation to optically thin. 
The
thickness of this layer is only around 500 km and the temperature ranges from
around 6200\,K at Rosseland mean optical depth unity to close to 4000\,K at the
uppermost parts.
The photosphere is the boundary between the solar interior and the atmosphere, and for this article I assign it to the interior. This is principally because the measurement of element abundances in the photosphere is particularly relevant to solar interior studies (Section~\ref{sec:abun}).

Interest in the interior has a strong overlap with Astrophysics, where the Sun is a reference object for topics such as stellar structure, stellar dynamos and element composition in the universe. In contrast to the highly dynamic solar atmosphere, the solar interior is mostly viewed as a constant except for the rhythm of the 11-year solar cycle.
The physics of the solar interior are well-established to the extent that Standard Solar Models (SSMs) exist that yield excellent agreement with observed quantities, with the notable exception of the ``solar problem" described later.

Observationally, the interior is mostly studied from ground-based observatories. When viewed at high spatial and temporal resolution, the photosphere shows continuous convective motions organized into granules and supergranules of sizes around 1 and 30\,Mm, respectively. Helioseismology is the measurement and interpretation of oscillations of the solar surface, usually obtained from Doppler shifts of visible absorption lines. Wave modes of different frequencies penetrate to different depths in the interior hence helioseismology can yield quantities such as pressure, density and sound speed in the interior---crucial for testing SSMs.

There are numerous ground-based solar telescopes around the world. Large telescopes such as the Goode Solar Telescope (California), the Daniel K.\ Inouye Solar Telescope (Hawaii), the Swedish Solar Telescope and the GREGOR telescope (both in the Canary Islands) yield  images of the photosphere down to a resolution of around 0.1\as\ (70\,km). The  Global Oscillation Network Group (GONG) is a network of six smaller-scale telescopes distributed so as to give almost 24-hour coverage of the Sun. Each telescope observes the entire solar disk and the data are used for helioseismology. Space-based visible telescopes can yield 24-hour coverage of the Sun and are free from distortion effects due to atmospheric seeing, however the sizes of the telescopes are limited. The most important space instrument is the SDO/HMI, which obtains full disk Dopplergrams for helioseismology and magnetograms for studies of the solar magnetic field.

In the 21st Century a great deal of attention has been paid to what has been referred to as the ``solar abundance problem" or the ``solar modeling problem" (here simply ``solar problem" is preferred). As discussed in Section~\ref{sec:abun}, element abundances are derived from modeling of solar photospheric spectra and used as input to SSMs. Advanced 3D models have replaced earlier 1D models and yielded abundances that can be significantly different, depending on the analysis method and lines studied. SSMs that previously gave results that agreed well with parameters derived from helioseismology no longer do so  with some of  the newer abundance results \cite{2008PhR...457..217B}. For example,  predictions of sound speed profiles, the depth of the convection zone, and the helium abundance in the solar envelope are all modified when the photospheric abundances change.
Atomic data are highly relevant to this problem, as discussed in the following section.

\section{Applications of Atomic Data to the Solar Interior and Photosphere}\label{sec:int-appl}

There are three main areas to which atomic data are important for the solar interior. Nuclear cross-section rates and atomic opacities are critical for SSMs, while measurements of element abundances in the photosphere depend on atomic cross-sections and oscillator strengths.

\subsection{The solar core: nuclear reactions}\label{sec:core}

\begin{figure}
    \centering
    \includegraphics[width=0.45\textwidth]{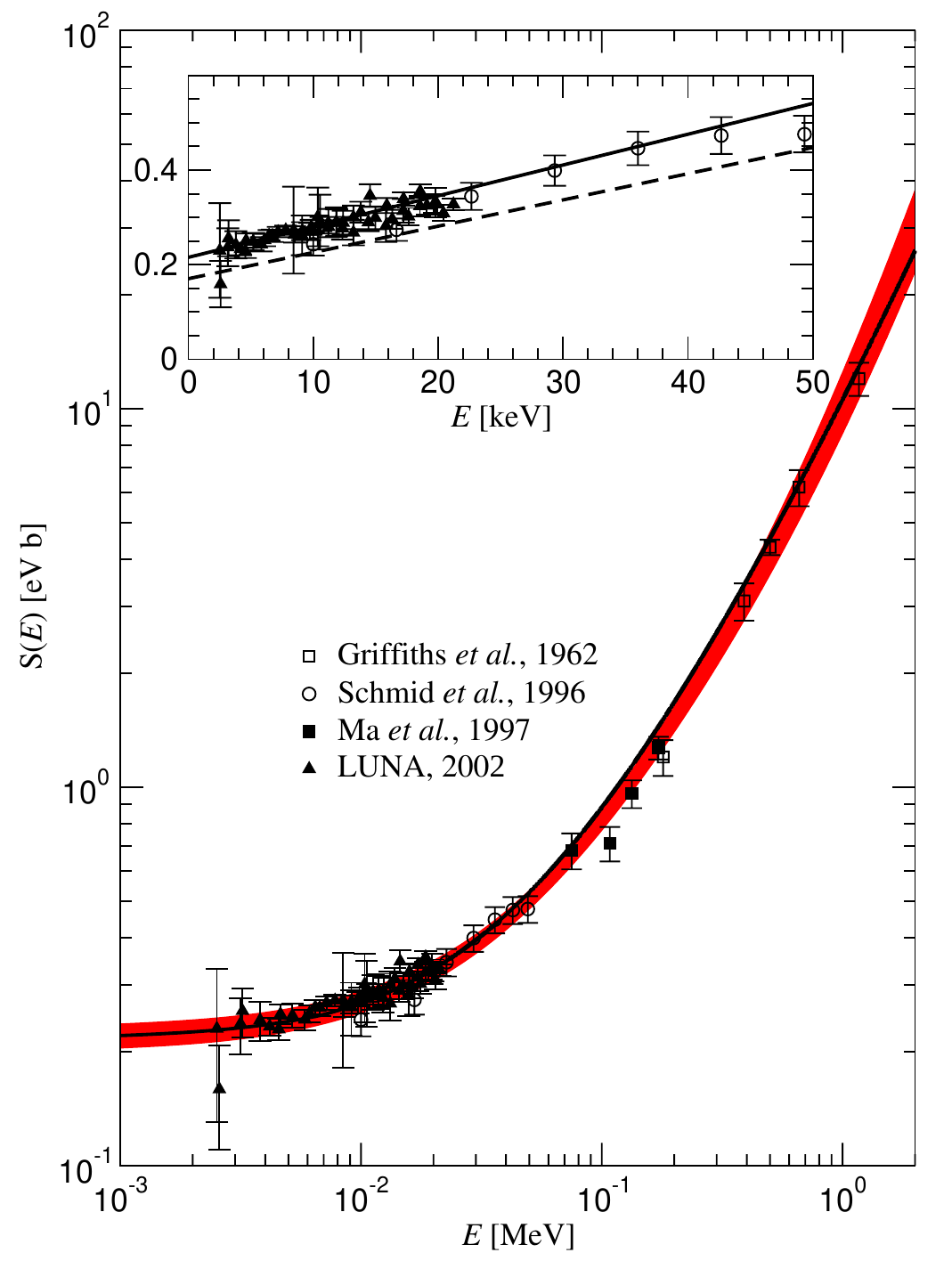}
    \caption{A comparison of four experimental measurements of $S$ for the $^2$H + p $\rightarrow$ $^3$He + $\gamma$ reaction with a theoretical calculation (black line). The broad red line shows the fit to the experimental data. The inset planel shows a close-up of the low energy points. From \cite{2011RvMP...83..195A}.}
    \label{fig:adelberger}
\end{figure}

Nuclear reactions in the Sun's core provide the energy source for the Sun's radiative emissions, and the dominant process is the proton--proton (pp) chain that produces $^4$He. A combination of experimental and theoretical cross-sections are required for various reactions that comprise the pp chain and its sub-branches. For example, the initial reaction that fuses two protons to yield deuteron has a cross-section that is too small to be measured in the laboratory. For reaction cross-sections that can be measured in the laboratory, the energy range is often higher than the energies relevant in the Sun and hence theory is used to extrapolate to lower energies. Measurements and theory yield the $S$-factor, which is derived from the cross-section by removing some of the rapidly-varying energy terms.  Figure~\ref{fig:adelberger} shows  the $S$-factor for the $^2$H + p $\rightarrow$ $^3$He + $\gamma$ reaction. Agreement between the theoretical calculation (solid line) and the measured values is excellent.
Comprehensive reviews of the nuclear cross-sections that recommended best values and uncertainties were published in 1998 and 2011 \cite{1998RvMP...70.1265A,2011RvMP...83..195A}, and a further update has recently been prepared \cite{2024arXiv240506470A}.

Models for the nuclear processes can be checked indirectly through their input into Standard Solar Models followed by comparisons with helioseismology, and directly through their prediction of neutrino fluxes. The ``solar neutrino problem," whereby the flux of electron neutrinos at the Earth was a factor three lower than expected was resolved in the early 2000s following measurements by Super-Kamiokande \citep{2001PhRvL..86.5651F} and the Sudbury Neutrino Observatory \citep{2001PhRvL..87g1301A} that demonstrated that electron neutrinos change their flavor during their passage to the Earth, explaining the lower fluxes.

In the push for improved accuracy for neutrino measurements, underground experiments have become a priority  in order to reduce contamination by cosmic rays. An example is the Laboratory for Underground Nuclear Astrophysics (LUNA) at Gran Sasso, Italy, which specializes in low-energy reactions for Astrophysics (Figure~\ref{fig:adelberger}).

\subsection{Opacity}\label{sec:opacity}

M.~Schwarzschild \cite{1958ses..book.....S} stated in 1958: ``For the astronomer who tries to reconstruct the stellar interior the opacity is by far the most bothersome factor in the entire theory." and it continues to be a major focus for solar scientists and atomic physicists. 
Opacity is the degree to which radiation is absorbed by a plasma and it comprises the individual processes photoexcitation (bound-bound), photoionization (bound-free), inverse bremsstrahlung, and photon (Thomson) scattering.  
In the past, stellar interior codes used as input the Rosseland mean opacity, which is the total opacity integrated over frequency.  Modern codes such as Modules for Experiments in Stellar Astrophysics \citep{2011ApJS..192....3P}, CESAM2K \citep{2008Ap&SS.316...61M} and the Toulouse-Geneva Evolution Code \cite{2012A&A...546A.100T} directly use opacity tables from, e.g., the Opacity Project \cite{1987JPhB...20.6363S} or OPAL \cite{1996ApJ...464..943I}.

The opacity has a dependence on both element abundances and temperature, such that an element's contribution changes with location within the interior. The region close to the convective/radiative zone boundary is of particular interest as the opacity here effectively determines where the boundary occurs. The boundary has been accurately measured from helioseismology \cite{1991ApJ...378..413C} giving a good point of comparison with SSMs.  The largest contributors to opacity near the boundary are oxygen, iron and neon \cite{2009PhPl...16e8101B}. 
The oxygen opacity remains uncertain due to controversy over the photospheric oxygen abundance. Recent values are generally smaller than the standard oxygen abundance from the 1990s \cite{1993oee..conf...15G,1998SSRv...85..161G}, with one value up to 34\%\ lower \cite{2021A&A...653A.141A}, while another is 17\%\ lower \cite{2021MNRAS.508.2236B} but consistent  within the uncertainties.
Neon can not be directly measured in the photosphere \cite{2021A&A...653A.141A}, but can be measured relative to oxygen in the solar atmosphere \cite{2018ApJ...855...15Y}. Hence a lower oxygen abundance indirectly lowers the neon abundance, too, further reducing the opacity. The result is a significant reduction in the convection zone depth in SSMs, giving a large discrepancy with the empirical value \cite{2004PhRvL..92l1301B}.

The opacity data used in SSMs are largely from theoretical calculations, and the discrepancy could be resolved if the theoretical data underestimate actual opacities by 15--20\%. The main sources of opacities are 
the Opacity Project  and OPAL, but good agreement is found between the two with no evidence of errors that could explain the solar problem \cite{2005MNRAS.360..458B}.

\begin{figure}
    \centering
    \includegraphics[width=0.45\textwidth]{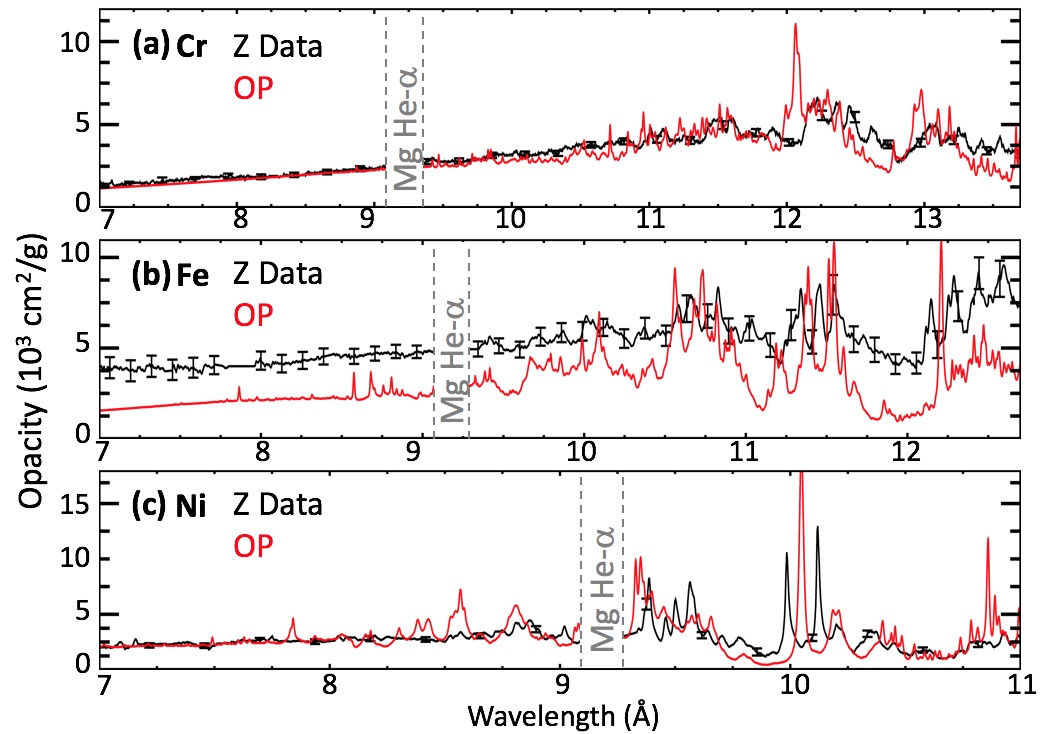}
    \caption{Comparisons of model (red line) and measured (black line) opacities for Cr, Fe and Ni at a temperature of 180\,eV and a density of $3\times 10^{22}$\,cm$^{-3}$.  Significant differences are found for Fe, but not Cr and Ni. From \cite{2019PhRvL.122w5001N}.}
    \label{fig:nagayama}
\end{figure}

Opacities have been measured in the laboratory since the 1980s \cite{1988ApPhL..52..847D}, and recent work has focused on iron opacities in the energy region corresponding to the solar tachocline. In a surprising result \cite{2015Natur.517...56B}, the measured opacities for iron (charge states 16+, 17+ and 18+) were found to be higher than the theoretical values by amounts ranging from 30\%\ to 400\%, depending on wavelength. A follow-up work \cite{2019PhRvL.122w5001N} on Cr and Ni opacities found significantly smaller discrepancies (Figure~\ref{fig:nagayama}), which suggests physics problems specifically for Fe at the temperatures and densities sampled by the experiment. New theoretical work has been performed for \ion{Fe}{xvii} \cite{2021MNRAS.508..421D} and \ion{Fe}{xvii--xix} \cite{2024JPhB...57l5002N,2024MNRAS.527L.179P} to investigate the source of the experimental discrepancy, but the opacities remain relatively unchanged.

\subsection{Photospheric abundances}\label{sec:abun}

Element abundances derived from the Sun's photospheric spectrum are important inputs to SSMs. 
They are derived in a two-step process whereby a 3D hydrodynamic model of  a small area of the solar surface that includes the sub-surface convection zone and photosphere is first created. Temporal snapshots are then extracted from the simulation and subject to radiative transfer models to predict the absorption profile of the line of interest. The emission in a line depends on the level populations of the atom or molecule, which can be assumed to be in local thermodynamic equilibrium (LTE). More commonly in modern calculations, non-LTE is assumed whereby the balance between multiple atomic processes is solved to yield the populations. 
Unlike earlier 1D models, the 3D models do not have any free parameters that can be adjusted and are found to accurately reproduce parameters such as the photospheric temperature \cite{2000A&A...359..729A}, giving confidence in their accuracy. The abundance itself is usually obtained by fitting the synthetic line profile to one or more solar spectra \cite{2008A&A...488.1031C}, but fits to the equivalent widths can also be performed.

The initial results from the 3D models generated controversy in the early 2000s, as the derived C, N and O abundances were significantly lower than the values previously used in Astrophysics. Due to the importance of these elements to SSMs (see above) the results have been revisited several times and generally been confirmed \cite{2021A&A...653A.141A}, although there remain differences between different authors \cite{2022A&A...661A.140M}.

\begin{figure}
    \centering
    \includegraphics[width=0.4\textwidth, trim=0cm 0cm 12cm 22cm, clip]{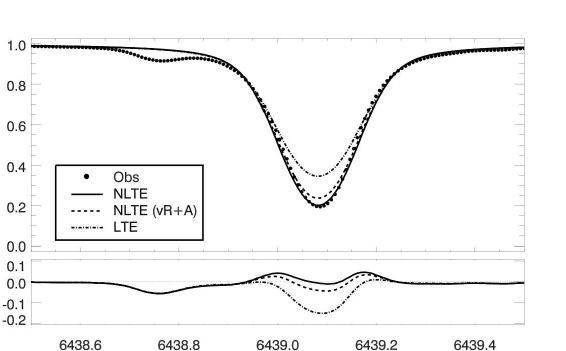}
    \caption{An example of modeling a photospheric absorption line to infer the element abundance. The line is \ion{Ca}{i} 6439\,\AA, and three modeled profiles are shown. The lower panel shows the differences between the observed profile and modeled profiles. From \cite{2019A&A...623A.103O}.}
    \label{fig:allende}
\end{figure}

Atomic data play a critical role in the abundance measurements in several ways. Most directly, accurate oscillator strengths are needed for the strengths of the absorption lines. To fit the profile shapes, Stark and van der Waals broadening parameters are needed. For the non-LTE level population calculations, cross-sections for electron and hydrogen collisions, and photoionization are also needed, together with accurate energy levels \cite{2016A&ARv..24....9B}.

The process of validating the abundances obtained by the 3D NLTE modeling is complex. Taking oxygen as an example, both atomic and molecular (OH, CO) transitions can be used and care has to be taken to account for blending both in the cores of the lines and the wings. Center-to-limb variation of the lines' observed profiles needs to be modeled depending on the observed spectra being fit (e.g., disk-center, near-limb, or disk-averaged), but can also be used to constrain atomic parameters \cite{2015A&A...583A..57S}. Due to the large differences with earlier 1D models, it is typical to compare the 3D model results with various flavors of 1D models. For example, averaging the 3D hydrodynamic model over space and time to create a 1D model, or using theoretical and semi-empirical models 1D models with the same equation of state, radiative transfer methods and opacities \cite[e.g.,][]{2015A&A...583A..57S,2021A&A...653A.141A}.

Figure~\ref{fig:allende} \cite[from][]{2019A&A...623A.103O} illustrates the change in an absorption line profile   from LTE to non-LTE, and also the impact of using accurate atomic data (the NLTE curve) versus simple approximations to the data (the vR+A curve, where ``vR" and ``A" refer to standard formulae for electron excitation and ionization rates due to van Regemorter and Allen). The example is the \ion{Ca}{i} 6439\,\AA\ line, and the same calcium abundance and macroscopic broadening parameters are used for the three curves. In this case the accurate atomic data do not have a significant impact on the derived abundance, but do lead to an improved fit to the line profile.

\section{The Solar Atmosphere}\label{sec:atmosphere}

\begin{figure*}[t]
    \centering
    \includegraphics[width=\textwidth]{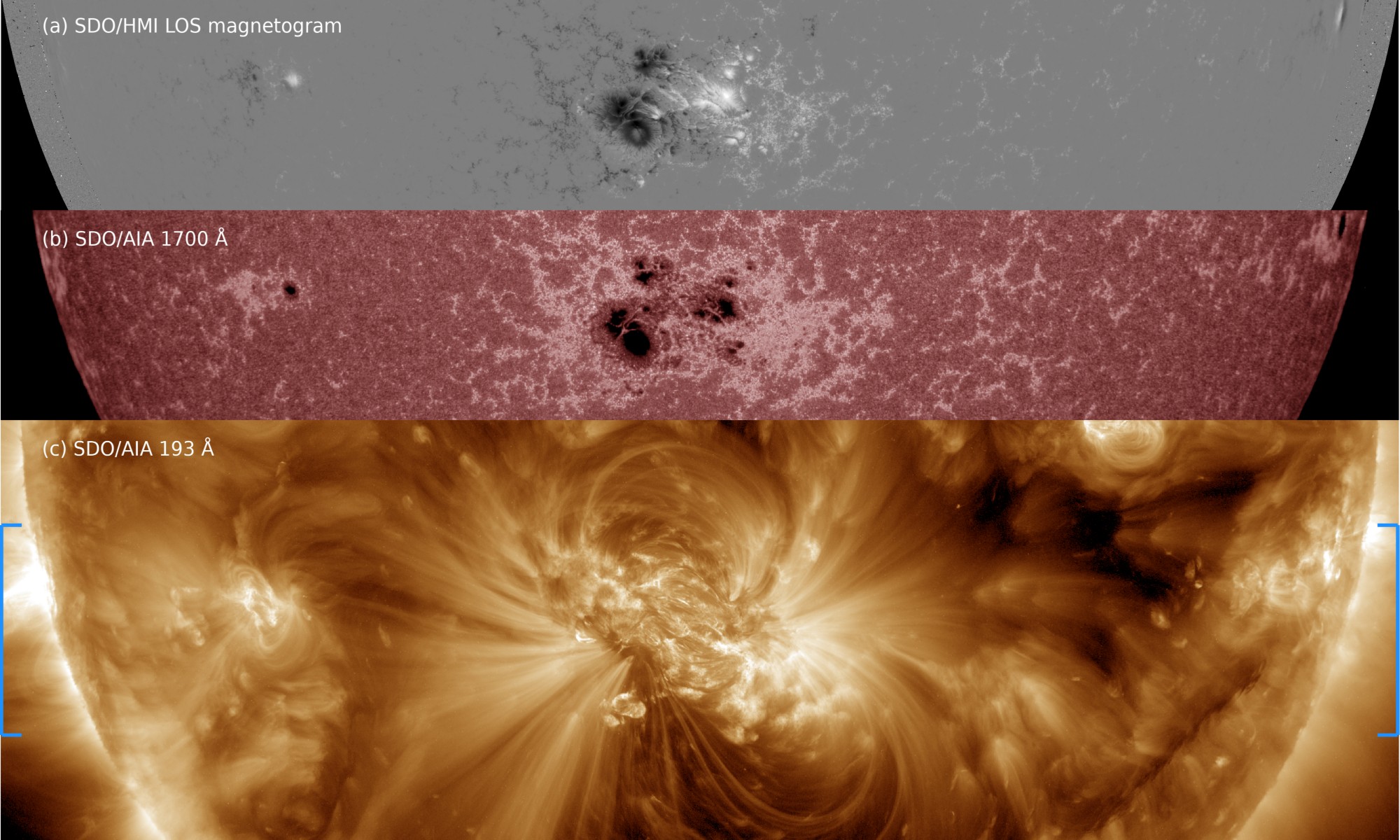}
    \caption{Images of the Sun from 23 October 2014 at 12:00~UT. Panel (a) shows the photospheric  LOS magnetic field from SDO/HMI, scaled within $\pm$~2000\,G; Panel (b) shows a chromospheric continuum image at 1700\,\AA\ from SDO/AIA; and Panel (c) shows a coronal emission line image at 193\,\AA, also from SDO/AIA. A logarithmic intensity scaling is used for the latter two Panels. The blue lines on Panel (c) indicate the sub-regions shown in Panels (a) and (b).} 
    \label{fig:sdo}
\end{figure*}

The Sun's atmosphere is strikingly revealed during total solar eclipses, where lobes and plumes of emission can be seen extending 2--3\,$R_\odot$ from the solar limb. The emission is principally photospheric light scattered by free electrons in the Sun's corona. The high temperature of the corona (1--2\,MK) was established in the 1940s and implied that the bulk of  the corona's emission should be at extreme ultraviolet (EUV) and X-ray wavelengths. Multiple spacecraft now routinely image the corona at EUV wavelengths, including NASA's flagship  SDO, which obtains full-disk images in seven EUV wavelengths at 12\,s cadence and 1.2\as\ (900\,km) spatial resolution. Figure~\ref{fig:sdo}(c) shows a section of an SDO/AIA image obtained at a wavelength of 193\,\AA.

The complex structure of the corona seen in this image is driven by the convective motions in the subsurface layers, which stretch, squeeze, twist and move the magnetic field that passes through the photosphere. The low density of the corona means the plasma is constrained to move along magnetic field lines, giving loops and plumes of emission according to if the magnetic field is closed or open. The 3D coronal magnetic field can not be routinely measured, but estimates are possible through coronal seismology \cite{2001A&A...372L..53N}, radio gyroresonance emission \cite{2020Sci...367..278F} and polarization of coronal forbidden lines \cite{2000ApJ...541L..83L}. Instead, the coronal field is usually extrapolated from the photospheric field, which is routinely measured by SDO/HMI. Figure~\ref{fig:sdo}(a) shows the line-of-sight (LOS) magnetic field strength in the photosphere, as measured by SDO, with white and black denoting strong fields of opposite polarity and gray denoting areas of weak magnetic field. The solar surface is uniformly covered with a network of  small-scale, weak magnetic field structures that are organized by the sub-surface convective motions. Large-scale strong field is associated with active regions, such as the one in the center of Figure~\ref{fig:sdo}(c). These generally have a bipolar structure. 

The temperature of the photosphere falls with height to a minimum of 4000\,K, marking where the chromosphere begins. This is a region where temperature increases to 10,000\,K at heights of around 10\,Mm and it
can be studied through  lines in the visible (particularly H$\alpha$ and \ion{Ca}{ii} H \& K) and UV (\ion{Mg}{ii} h \& k). The UV continuum is also important for the chromosphere, and Figure~\ref{fig:sdo}(b) shows an image from the SDO/AIA 1700\,\AA\ channel. 
Comparing with the upper panel, the dark sunspots can be identified with the most intense magnetic field. Intermediate magnetic field strengths correlate with areas of brightness in the chromosphere that are termed plage. The temperature rises sharply from the chromosphere to the corona through a thin layer called the transition region: the large temperature gradient is due to the large conductive flux from the hot corona.

The wavelength region 150--1600\,\AA\ is particularly important for observing the solar atmosphere as it features emission lines from the chromosphere to the corona. Transition region lines are mostly found above an approximate dividing line at 400\,\AA, and coronal lines below this wavelength.  
A group of strong resonance lines of \ion{Fe}{ix--xiv} between 170--212\,\AA\ are particularly important. Interest in the far-UV and EUV has motivated a wide range of space instrument designs and current and future concepts are discussed in \cite{2021FrASS...8...50Y}.

\section{Solar Atmosphere Atomic Data Requirements}\label{sec:atm-appl}

The low density of the corona (around $10^9$\,cm$^{-3}$) means the plasma is not in thermodynamic equilibrium and so it is necessary to model the atomic processes that excite and de-excite atomic levels. The most important processes are electron excitation and spontaneous radiative decay. Depending on the ion, transition data for up to 1000 levels may be required to yield accurate emissivities for transitions observed in the solar spectrum. Over 200 ions give rise to measurable lines in observed spectra and thus a huge amount of atomic data are required to accurately model coronal spectra. The CHIANTI atomic database has provided these data to the community for almost 30 years and is described in Section~\ref{sec:chianti}. Modeling the balance between ionization states is also crucial for modeling the coronal spectrum, and electron collisional ionization and recombination (both radiative and dielectronic) are the key processes. Timescales are significantly longer than for electron excitation and so level balance can be treated separately from ion balance, greatly simplifying the problem. In addition, total rates out of the ground state are usually sufficient. These rates are also provided in CHIANTI.

In the transition region and chromosphere, additional atomic processes can be important for low charge species (typically $3+$ or less). For example, metastable levels can have populations comparable to the ground state and so level-resolved ionization and recombination rates may be needed. Charge transfer, i.e., the exchange of an electron between the ion and H or He, and photoionization can compete with electron ionization and recombination. Many strong chromospheric lines are optically thick and so radiative transfer calculations are necessary for modeling. 
Modeling of solar atmosphere structures that fully takes into account matter--radiation interactions is currently only possible in 1D \cite{2020LRSP...17....3L}, an  example being the  RADYN code \cite{2015ApJ...809..104A} that is widely used for solar flare modeling. 3D models of solar atmosphere structures need to include the magnetic field, and so magnetohydrodynamic simulations are performed. Examples include the Bifrost \cite{2011A&A...531A.154G} and MURaM \cite{2005A&A...429..335V} codes.
Fully including the effects of radiative transfer on the plasma is beyond current computational capabilities, and so simple recipes are employed such as lookup tables \cite{2012A&A...539A..39C}.

Examples of recent work in applying atomic physics to solar atmosphere data are discussed below, and the CHIANTI atomic database is described.

\subsection{The CHIANTI atomic database}\label{sec:chianti}

CHIANTI was named for the wine-growing region in Tuscany near Florence where a number of meetings were held in the formative years of the project. The first version was released in 1996 \cite{1997A&AS..125..149D} and, unusually for the time, it was completely open source. CHIANTI consists of a database of atomic parameters for ions, plus software packages written in IDL and Python that take these data and allow the user to compute the radiative emission from a plasma.

\begin{figure*}
    \centering
    \includegraphics[width=\textwidth]{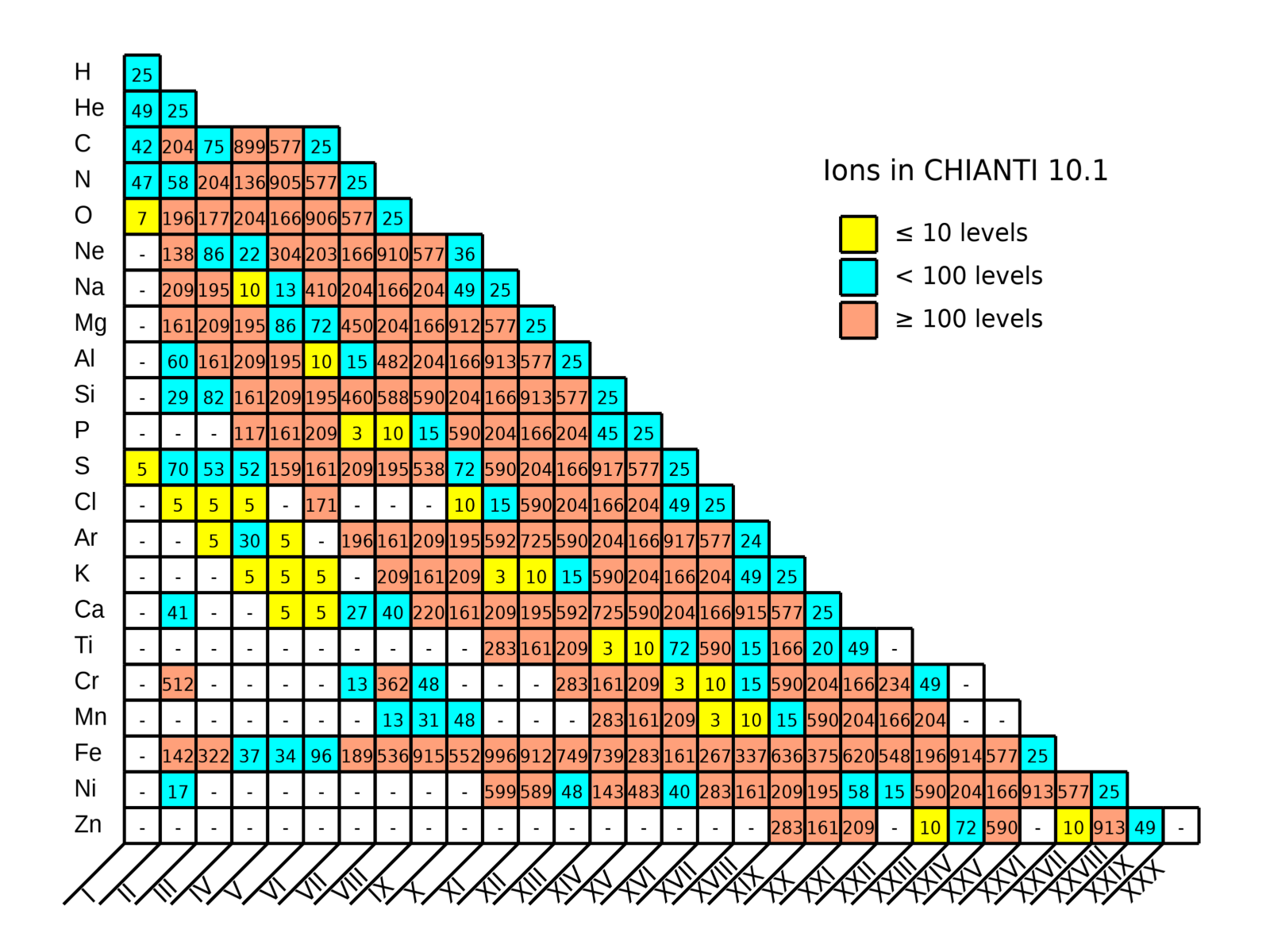}
    \caption{A table showing the ions that are modeled in the CHIANTI atomic database. The number in each box indicates the number of fine-structure levels of the ion model.}
    \label{fig:chianti}
\end{figure*}

The core data holdings are electron excitation rates, spontaneous radiative decay rates and level energies. Only fine structure ($J$-resolved) levels are included in order to obtain accurate transition wavelengths. Figure~\ref{fig:chianti} shows ions for which CHIANTI has data, and the number of levels for each ion is indicated. 

A crucial part of CHIANTI has been the assessment of electron collision strengths through a graphical procedure that follows that recommended by \cite{1992A&A...254..436B}. This has been invaluable for identifying problematic data, and the team has frequently communicated with atomic physicists to understand and resolve discrepancies. The team members are also practising spectroscopists who use CHIANTI for their research, and they have performed many benchmark analyzes to assess the accuracy of the database, e.g., \cite{1998A&A...329..291Y, 2002ApJS..139..281L,2012A&A...537A..38D}.

As can be seen from Figure~\ref{fig:chianti} there remain many ions for which CHIANTI does not have data. These are either because the ion does not yield significant lines in the solar spectrum, or because fine-structure electron excitation data do not exist for the ion. 

At UV and EUV wavelengths emission lines are mostly resolved and so high accuracy data are needed for specific emission lines. The lines generally come from $n=2,3$ states, and so model atoms do not need to be very large to model the UV lines.
At X-ray wavelengths, however, lines typically can come from $n=4, 5, 6$ states, so  much larger atomic models are needed. Since line density at X-ray wavelengths is high and solar X-ray spectrometers generally have low spectral resolution then there is less requirement for high accuracy data, however.

\subsection{Line identifications: Fe\,VII and Fe\,IX}\label{sec:fe7}

\begin{figure*}
    \centering
    \includegraphics[width=\textwidth]{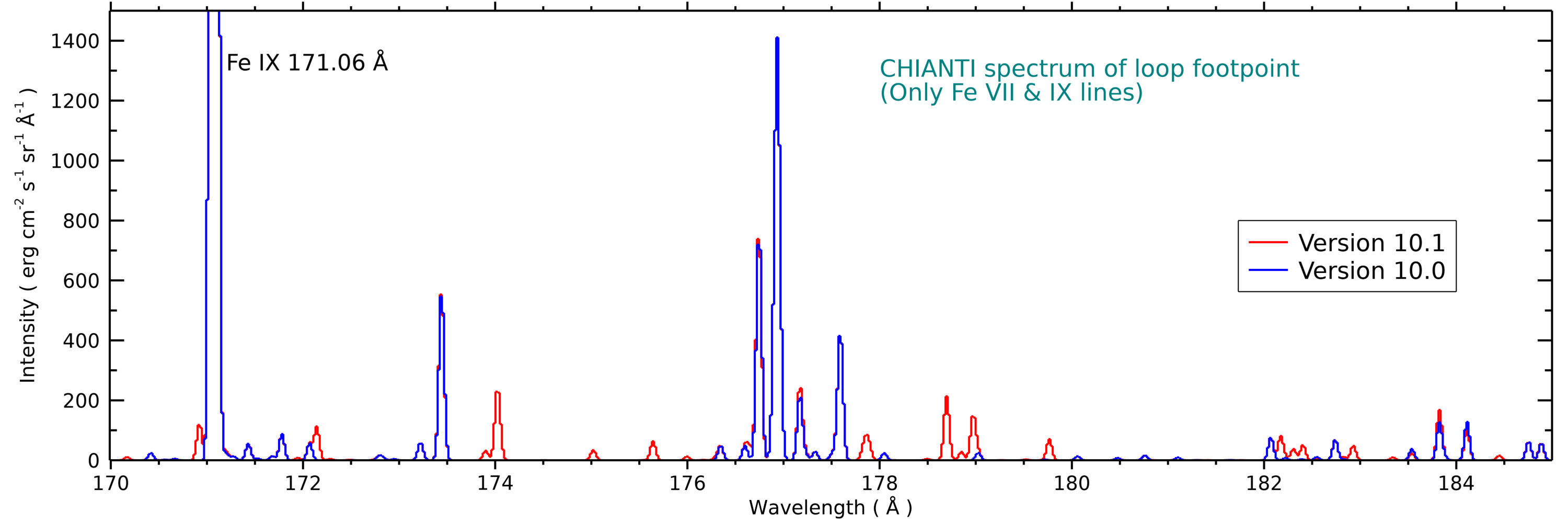}
    \caption{CHIANTI synthetic spectrum in the 170--185\,\AA\ region showing only \ion{Fe}{vii} and \ion{Fe}{ix} lines. The blue line shows the spectrum from CHIANTI 10.0 and the red lines shows new lines that were added in CHIANTI 10.1. The spectra were generated with the  differential emission measure curve of \cite{2009ApJ...706....1L} and a density of $10^9$\,cm$^{-3}$.}
    \label{fig:fe7}
\end{figure*}

\ion{Fe}{vii} and \ion{Fe}{ix} yield hundreds of weak emission lines in the 150--300\,\AA\ region that result from $3p^53d^3$ and $3p^43d^2$ configurations, respectively. These configurations have 110 and 111 fine structure levels that exhibit very significant level mixing making line identification difficult. \cite{2021ApJ...908..104Y,2022ApJS..258...37K,2022ApJ...936...60R} have utilized solar EUV spectra from Hinode/EIS and high-resolution laboratory spectra to perform new line identifications for these ions. Figure~\ref{fig:fe7} shows some of the new identifications in the 170--185\,\AA\ wavelength region. A CHIANTI synthetic spectrum is displayed that contains only lines from the two ions. The red lines are the new identifications that were added to CHIANTI 10.1 \cite{2023ApJS..268...52D}. The displayed region contains the very strong \ion{Fe}{ix} 171\,\AA\ line, and the new data will be relevant to the upcoming MUSE and Solar-C missions (Table~\ref{tbl:missions}) which will observe this wavelength range. 

Despite the progress for these two ions, 51\%\ and 82\%\ of the $3p^53d^3$ and $3p^43d^2$ levels do not have experimental energies. To make further progress more accurate electron scattering calculations need to be performed for both ions in order to yield more accurate predicted line emissivities. In addition, new high resolution laboratory spectra over a wide wavelength band (e.g., 100\,\AA) are required that can discriminate between neighboring charge states. A wide wavelength region is crucial for identifying Ritz combinations, as done for \ion{Fe}{ix} \cite{2022ApJ...936...60R}.

\subsection{Experimental energies: Mg\,VII and Si\,VII}\label{sec:energies}

The National Institute for Standards and Technology (NIST) maintains the Atomic Spectra Database (ASD) \cite{nist} that is the standard reference for ion wavelengths. It is compiled from laboratory and astrophysical wavelength measurements that are combined to yield an optimized set of energy levels from which Ritz wavelengths are calculated. An alternative set of reference wavelengths is provided by B.~Edl\'en in a series of papers between 1983 and 1985 for isoelectronic sequences from Be through F. Unlike the NIST compilation, experimental wavelength measurements are supplemented by theoretical calculations and extrapolations along the sequences. The Edl\'en work is also limited to $n=2$ configurations.

\begin{table}[h]
\caption{Reference wavelength comparison for \ion{Si}{vii} and \ion{Mg}{vii}.}\label{tbl:mg7}%
\begin{tabular}{@{}ccccc@{}}
\toprule
& & \multicolumn{2}{c}{$\lambda$ (\AA)} \\
\cmidrule{3-4}
Ion & Transition & ASD & Edl\'en & $\Delta\lambda$ (\kms)\\
\midrule
\ion{Si}{vii}\footnotemark[1] & $^3P_2$--$^3P_2$ & 275.353 & 275.361 &  9 \\
 & $^3P_0$--$^3P_1$ & 276.839 & 276.850 & 12 \\
\ion{Mg}{vii}\footnotemark[2] & $^3P_0$--$^3S_1$ & 276.154 & 276.138 & 17 \\
& $^1D_2$--$^1P_1$ & 280.737 & 280.722 & 16 \\
\botrule
\end{tabular}
\footnotetext[1]{Configuations: $2s^22p^4$--$2s2p^5$.}
\footnotetext[2]{Configuations: $2s^22p^2$--$2s2p^3$.}
\end{table}

The CHIANTI team has made use of both NIST and Edl\'en data, and have found significant differences in some cases. \cite{2023ApJ...958...40Y}  highlighted problems for C-like \ion{Mg}{vii} and O-like \ion{Si}{vii} and examples are given in Table~\ref{tbl:mg7}. These lines are all observed by Hinode/EIS, which can measure wavelengths to a precision of 1\,\kms\ and an accuracy of 5\,\kms. The wavelength differences between the ASD and Edl\'en wavelengths are thus very significant. 

By combining wavelengths measured from Hinode/EIS with forbidden transition wavelengths from Astrophysical nebula observations (which were unavailable at the time of the NIST and Edl\'en compilations), a new set of reference wavelengths could be generated for the two ions \cite{2023ApJ...958...40Y}. Better agreement was found with the Edl\'en values than the NIST values.

This example demonstrates the importance of keeping the NIST database up to date with the latest developments in space-based wavelength measurements. This requires continued funding for NIST as well as active collaboration of astrophysicists with the NIST team.

\subsection{Improved ionization balance modeling}\label{sec:ioniz}

CHIANTI models the ionization balance for an element through a balance between electron collisional ionization and recombination, as first introduced in CHIANTI 6 \cite{2009A&A...498..915D}. An update was performed in CHIANTI 10 \cite{2021ApJ...909...38D} to model the suppression of dielectronic recombination (DR) at high densities. In a series of papers \cite{2019A&A...626A.123D,2020MNRAS.497.1443D,2021MNRAS.503.1976D,2021MNRAS.505.3968D}, R. Dufresne and colleagues have developed advanced atomic models for several isonuclear sequences that incorporate additional effects that impact the ionization balance, particularly for low charge states in the chromosphere and low temperature transition region. 

\begin{figure}
    \centering
    \includegraphics[width=0.45\textwidth]{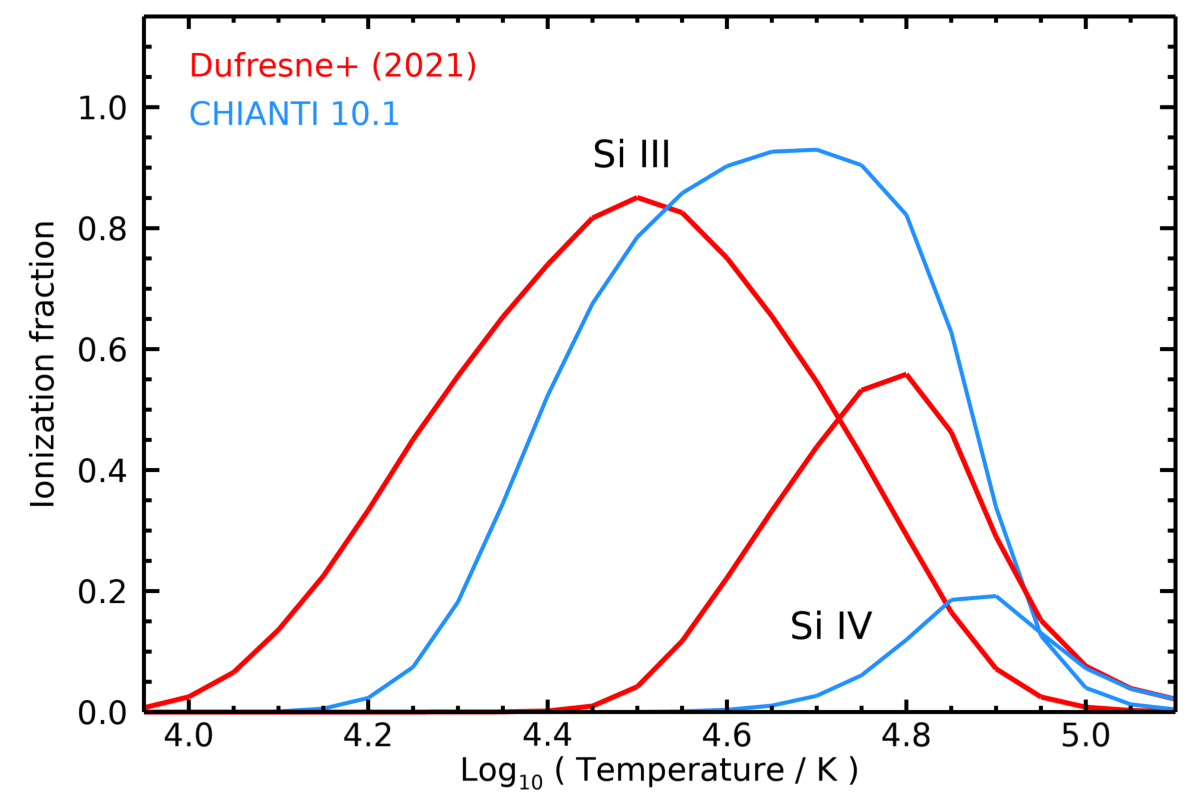}
    \caption{Ion fractions for \ion{Si}{iii} and \ion{Si}{iv} computed from the Dufresne advanced models (red) and CHIANTI 10.1 (blue).}
    \label{fig:dufresne}
\end{figure}

Figure~\ref{fig:dufresne} compares ion fractions for \ion{Si}{iii} and \ion{Si}{iv} from the advanced models \cite{2021MNRAS.505.3968D} with those from CHIANTI 10.1 \cite{2023ApJS..268...52D}. The advanced models include density effects (level-resolved ionization/recombination and DR suppression), photoionization and charge transfer, and a constant pressure of $3.0\times 10^{14}$\,K\,cm$^{-3}$ was assumed.
 
CHIANTI 11 is scheduled to be released in 2024 and will include modified versions of the Dufresne advanced models for the elements C, N, O, Ne, Mg, Si and S. The models will incorporate the density effects of level-resolved ionization/recombination, DR suppression and charge transfer on the ionization balance for these elements. While the new models represent a significant advance over the previous CHIANTI modeling, three areas can be identified where new atomic calculations would be valuable.

\begin{enumerate}
    \item \textit{DR suppression}. The method for calculating this dates back to 1974 \cite{1974MNRAS.169..663S}, where suppression factors as a function of density were computed. The more recent work of \cite{2018ApJS..237...41N} applied  these factors to recent DR rates. Given the importance of DR suppression \cite{2018ApJ...857....5Y,2019A&A...626A.123D} in shifting transition region ions to lower temperatures, a modern calculation of the suppression factors is highly desirable. 
    \item \textit{Level-resolved ionization rates}. Ionization rates from the ground states of ions can be calculated and ``calibrated" against laboratory-measured rates in many cases \cite{2007A&A...466..771D}. For the advanced models discussed above, rates from metastables are required that are generally not available in the literature. An empirical formula \cite{1983MNRAS.203.1269B} can be used to scale the metastable rates to the ground rates \cite{2021MNRAS.505.3968D}. Systematic ab initio calculations along isoelectronic sequences of metastable rates are desirable, together with laboratory measurements for comparison.
    \item \textit{Recombination rates}. N.~Badnell and colleagues \cite{2003A&A...406.1151B,2006ApJS..167..334B,2018A&A...610A..41K} have systematically produced radiative and dielectronic recombination rates for isoelectronic sequences that are used in CHIANTI and other plasma codes. The data are available for all isoelectronic sequences from H to Si, plus the Ar sequence. Data for the P, S and Cl sequences are needed, and for sequences of the fourth row, particularly for iron ions.
\end{enumerate}

\section{Summary}\label{sec:summary}

A number of topics in Solar Physics have been highlighted where atomic physics data are important. A division has been made between the solar interior (Section~\ref{sec:interior}) and the solar atmosphere (Section~\ref{sec:atmosphere}). Although Solar Physics is a mature field, the Sun remains an important plasma laboratory that can be studied in great detail and provides a testing ground for Astrophysics and plasma theories. High quality atomic data will therefore continue to be required in the future, both theoretical calculations and laboratory measurements.

For further details and background on some of the topics covered in this article the following articles and books are recommended. The book of Pradhan \& Nahar \cite{2011aas..book.....P} details the atomic processes important to Astrophysics and how they are used to interpret spectra.  The review of Barklem \cite{2016A&ARv..24....9B} discusses the role of atomic physics in measuring the abundances of late-type stars, which is relevant to the Sun. The review of Adelberger et al.\ \cite{2011RvMP...83..195A}  gives a detailed summary of nuclear rates, including those relevant to the Sun. A short review of how various physical processess, including opacity and abundances, impact SSMs is given by Serenelli \cite{2016EPJA...52...78S}, and Asplund et al.\ \cite{2021A&A...653A.141A} give an overview of how 3D NLTE models are used to obtain photospheric abundances. Atomic data and plasma diagnostics relevant to the solar atmosphere are extensively covered in the book of Phillips et al.\ \cite{2008uxss.book.....P} and the review article of Del Zanna \& Mason \cite{2018LRSP...15....5D}.

\section{Declarations}

\subsection{Competing interests}

\subsection{Data availability}

There are no associated data with this article.

\subsection{Author contribution}

\backmatter

\bmhead{Acknowledgements}

The author thanks the referees for valuable comments that have improved the manuscript. Funding for this work was provided by  the NASA Heliophysics Data Resource Library, the GSFC Internal  Scientist Funding Model competitive work package program, and the Hinode project.

\bibliography{sn-bibliography}{}

\appendix

\section{Acronyms}

Table~\ref{tbl:acronyms} gives the acronyms used in this article.

\begin{table*}[t]
\caption{Acronyms used in this article.}\label{tbl:acronyms}%
\begin{tabular}{@{}ll@{}}
\toprule
SOHO & Solar and Heliospheric Observatory \\
STEREO & Solar Terrestrial Relations Observatory \\
SDO & Solar Dynamics Observatory \\
HMI & Helioseismic and Magnetic Imager \\
AIA & Atmospheric Imaging Assembly \\
EIS  & EUV Imaging Spectrometer \\
IRIS & Interface Region Imaging Spectrograph \\
CHASE & Chinese H-alpha Solar Explorer \\
ASO-S & Advanced Space-based Solar Observatory \\
PUNCH & Polarimeter to Unify the Corona and Heliosphere \\
MUSE & Multi-Slit Solar Explorer \\
\botrule
\end{tabular}
\end{table*}

\end{document}